# Testing the variants of the Stokes-Einstein relation in the framework of self-consistent generalized Langevin equation theory

Wanmei Zhang, Wenhan Zhang and Gan Ren[1]

[1]School of Science, Civil Aviation Flight University of China, Guanghan, China, 618307

**Abstract**

The two functional forms, $D \sim \tau^{-1}$ and $D \sim T/\tau$, are usually adopted as the variants of the Stokes-Einstein relation; where $D$ is the diffusion constant, $\tau$ the relaxation time and $T$ the temperature. The self-consistent generalized Langevin equation (SCGLE) theory is presented as an analytical tool to predict the long time dynamics of colloids and molecular liquids. In this work, taking truncated Lennard-Jones-like liquids as an example, the rationality of the two variants were tested in the framework of the SCGLE theory. Our results indicate that $D \sim \tau^{-1}$ is a good variant of the Stokes-Einstein relation in the framework of SCGLE theory; however, $D \sim T/\tau$ is not a good one but taking a fractional from as $D \sim (T/\tau)^\xi$ with an exponent $\xi \neq 1$ even the Stokes-Einstein relation is established in SCGLE theory.

## 1. Introduction

The Stokes-Einstein relation [1] $D = k_B T / \alpha$ relates the diffusion constant $D$ to the frictional coefficient $\alpha$ for a particle moving through a viscous fluid, where $k_B$ is the Boltzmann constant, $T$ is the temperature. The frictional coefficient is proportional to the shear viscosity $\eta$ and the effective hydrodynamic radius $a$, namely $\alpha = C\eta a$ [2], where $C$ is a constant determined by the boundary conditions. The Stokes-Einstein relation is a special case of fluctuation-dissipation theorem; which is satisfied in the linear response region and is breakdown out of equilibrium [1].

The Stokes-Einstein relation expressed as $D \sim T/\eta$ is observed to be invalid for liquid undergoes deep supercooling [3-8]. The dynamic properties of liquids show large changes as temperature decreases into the supercooled region. The rate of shear viscosity increases can be orders of magnitude lager than the diffusion constant decreases. It is difficult to accurately determine the shear viscosity in simulations. The alpha relaxation time $\tau$ is expected to have the same $T$ dependence as $\eta$, and which is usually adopted as a substitute of $\eta$. So it is proposed that the breakdown of the Stokes-Einstein relation is due to the decoupling of the diffusion and relaxation [7]. Two functional forms are usually adopted to evaluate the dependence of $\eta$ on $\tau$ in previous studies: $\eta \sim T\tau$ and $\eta \sim \tau$; the corresponding variants of the Stokes-Einstein relation are expressed as $D \sim \tau^{-1}$ [9-11] and $D \sim T/\tau$ [6, 12], respectively.

The variant $D \sim \tau^{-1}$ is based on the structural relaxation given by the self-intermediate scattering function defined by $F_s(k,t) = 1/N \sum_{j}^{N} \left\langle \exp\left\{ i\vec{k} \cdot \left[ \vec{r_j}(0) - \vec{r_j}(t) \right] \right\} \right\rangle$. If the displacement of particle $\delta \vec{r_j}(t) = \vec{r_j}(t) - \vec{r_j}(0)$ follows Gaussian, $F_s(k,t)$ can be expressed in terms of the mean square displacements of particle like $F_s(k,t) = \exp(-k^2 Dt)$ [7]. In simple liquids, $F_s(k,t)$ decays in an exponential function as $F_s(k,t) \sim e^{-t/\tau}$. Thereafter $D$ is coupled with $\tau$ like $D \sim \tau^{-1}$. The same functional form is also proposed in the mode coupling theory when the temperature is closed to the glass transition point [7]. The variant $D \sim T/\tau$ is based on the approximated relation $\eta = G_\infty \tau$ [12], where $G_\infty$ is the instantaneous shear modulus and is proposed to be a slight temperature dependent; $\tau$ here should be the stress relaxation time [2], but it is usually taken the structural time as a substitute in simulations. Two structural relaxation times, $\tau_s$ [2] and $\tau_n$ [6, 12], were adopted in variant $D \sim T/\tau$;

$\tau_s$ is given by $F_s(k,\tau_s) = e^{-1}$ and $\tau_n$ is given by $F(k,\tau_n) = e^{-1}$, where $F(k,t)$ is the coherent intermediate scattering function; $k$ is a wavevector and is usually chose the value corresponding to the first maximum of the static structure factor.

Although the two variants of the Stokes-Einstein relation were widely used to test the validity of Stokes-Einstein relation, their rationality is still in elusive. Shi et.al [2] have simulated three mixed Lennar-Jones-like liquids and coarse-grained ortho-terphenyl across a broad of temperatures and densities, aimed to investigate the rationality of the two variants by comparing with the results given by $D \sim T/\eta$. They found that $D \sim \tau^{-1}$ and $D \sim T/\tau$ will give different results comparing with $D \sim T/\eta$ as the temperature decreases. Although $D \sim T/\eta$ is still valid at certain temperature range, $D \sim \tau^{-1}$ and $D \sim T/\tau$ deviate; besides the deviations of $D \sim T/\tau$ are greater than $D \sim \tau^{-1}$. They proposed that one should critically evaluate the two variants while using to test the Stokes-Einstein relation.

It is very difficult to equilibrate the system at low temperatures to capture the dynamics; one needs to avoid the crystallization and simulate enough long time. In this work, we adopted the self-consistent generalized Langevin equation (SCGLE) theory proposed by Magdaleno et, al [13, 14] to test the rationality of the two variants of the Stokes-Einstein relation. The SCGLE theory is presented as an analytical tool to predict the long time dynamics of colloids and molecular liquids, which has correctly predicted the glass transition in colloids, dynamic equivalence between the colloids and molecular liquids with the same interaction as well as the dynamic equivalence between soft sphere and hard sphere [15-18]. Moreover, it has been successfully used to many liquids with different interaction, including truncated Lennard-Jones-like liquids [16], Yukawa fluid [17], and simple power law liquid [19, 20], etc. The SCGLE theory is an equilibrium theory and the Stokes-Einstein relation is valid. Although the supercooled liquid is metastable, yet it is still in equilibrium. So we can consider it as an ideal case to use SCGLE theory to test the two variants of the Stokes-Einstein relation. An advantage of the SCGLE theory is that one can calculate the dynamics and diffusion constant with the static structure factor; moreover, the static structure factor for enough rigid sphere particles can be numerically calculated by the Percus-Yevick approximation [21, 22] complemented with the correction of Verlet and Weis [23]. Another advantage is due to the dynamic equivalence between the soft sphere particle and rigid particles predicted by SCGLE theory; we only need calculation with one certain interaction to get knowledge of the dynamics with different interactions. In this work, taking enough rigid truncated Lennard-Jones-like liquids, we numerically tested the rationality of $D \sim \tau^{-1}$ and $D \sim T/\tau$ with two systems in the framework of SCGLE theory, one has a large volume fraction with a glass transition while cooling, another has a small volume fraction without glass transition even cooling to enough low temperatures. The paper is organized as follows: in section 2, we give a brief review of the SCGLE theory; section 3 is the results and discussion; the last is our conclusion in section 4.

## 2. Outline of the SCGLE theory

The SCGLE theory is a self-consistent theory composed of a series equations combined the coherent intermediate scattering function $F(k,t)$, self-intermediate scattering function $F_s(k,t)$ with the memory functions $C(k,t)$ or $C_s(k,t)$. The theory is based on some exact expressions of $F(k,t)$, $F_s(k,t)$ and generalized Langevin equation as well as complemented by a number of physically or intuitively motivated approximations. The main equations of the SCGLE theory are listed below, and the details can be referred to the references [13, 16, 17, 24].

The dynamic properties of molecular liquids are described by $F(k,t)$, $F_s(k,t)$ and diffusion constant. $F(k,t)$ and $F_s(k,t)$ is correlated with the density fluctuation $\delta n(r,t)$ of the local density $n(r,t)$ of molecular particles around its bulk equilibrium value $n$. $F(k,t)$ is described by $F(k,t) = \langle \delta n(k,t) \delta n(-k,0) \rangle$, where $\delta n(k,t) = 1/N \sum_{j=1}^{N} \exp[ik \cdot r_j(t)]$ is the Fourier transformation of $\delta n(r,t)$, $r_j(t)$ is the position of $j$th particle at time $t$. The Laplace transformation form of $F(k,t)$ and $F_s(k,t)$ in the SCGLE theory are expressed as

$$F(k,z) = \frac{S(k)}{z + \dfrac{k^2 D_0 S^{-1}(k)}{1 + C(k,z)}} \quad (1)$$

$$F_s(k,z) = \frac{1}{z + \dfrac{k^2 D_0}{1 + C_s(k,z)}} \quad (2)$$

where $D_0$ is the short time diffusion coefficient describing the particles' motion between collisions, $C(k,z)$ and $C_s(k,z)$ are the Laplace transform of the memory functions $C(k,t)$ and $C_s(k,t)$, respectively. $C(k,t)$ is approximated to be equaled to $C_s(k,t)$ in terms of the first-order Vineyard approximation, namely, $C(k,t) = C_s(k,t)$. The short time diffusion coefficient $D_0$ is given by theory of gas kinetics for molecular liquids, namely

$$D_0 = \frac{3}{8} \left( \frac{k_B T}{\pi M} \right)^{1/2} \frac{1}{n \sigma_{col}^2} \quad (3)$$

where $\sigma_{col}$ is the collision diameter of the particles and is equal to the diameter for rigid sphere. The memory function $C_s(k,t)$ is approximated by interpolating between its short and long wavelength limit, which is connected with the normalized times-dependent friction function $\Delta \xi^*(t)$, namely

$$C_s(k,t) = \lambda(k) \Delta \xi^*(t) \quad (4)$$

where $\lambda(k) = 1/\left[1 + (k/k_c)^2\right]$, $k_c$ is an empirically cutoff wavevector, we chose $k_c$ as the position of the first minimum follows the main peak of static structure factor $S(k)$[16]. Based on the generalized Langevin equation, $\Delta \xi^*(t)$ can be expressed by

$$\Delta \xi^*(t) = \frac{D_0}{3(2\pi)^3 n} \int d\vec{k} \left[ \frac{k[S(k)] - 1}{S(k)} \right]^2 F(k,t) F_s(k,t) \quad (5)$$

Equations (1), (2) and (5) are the main equations of SCGLE theory, which gives a closure relation of $F(k,t), F_s(k,t)$ and $C_s(k,t)$; and can be solved iteratively after given the equilibrium $S(k)$. The diffusion constant $D$ corresponds to the long time diffusion constant $D_L$ in SCGLE theory and can be calculated by

$$D_L = D_0 \bigg/ \left[1 + \int_0^\infty dt \Delta \xi^*(t)\right] \quad (6)$$

The relaxation times, $\tau_s$ and $\tau_n$ are determined by $F_s(k,\tau_s) = e^{-1}$ and $F(k,\tau_n) = e^{-1}$, where the wavevector $k$ is chose as the main maximum of $S(k)$.

## 3. Results and discussion

In this work, the truncated Lennard-Jones-like molecular fluids were adopted to examine the rationality of the two variants ($D \sim \tau^{-1}$ and $D \sim T/\tau$) of the Stokes-Einstein relation. The truncated Lennard-Jones-like potential is

$$u^{(v)}(r) = \varepsilon \left[\left(\frac{\sigma}{r}\right)^{2v} - 2\left(\frac{\sigma}{r}\right)^v + 1\right] \quad (7)$$

for $0 < r < \sigma$ and vanishes for $r \geq \sigma$; where $\varepsilon$ is the energy parameter, $\sigma$ is the interaction length parameter, $v$ determines the softness of the particle and $v \to \infty$ corresponds to the rigid sphere. The temperature is in unit $\varepsilon/k_B$ and the length is in unit $\sigma$. We chose $v=20$ for further numerical calculation, which is enough rigid to calculate $S(k)$ by the Percus-Yevick approximation complemented with the correction of Verlet and Weis [16]. The blip function [16, 25] is adopted to calculate the effective rigid sphere diameter and the corresponding effective volume fraction $\phi$. The blip function is

$$\int d^3r \left[\exp\left(-\beta u^{(v)}(r)\right) - \exp\left(-\beta u^{(HS)}(r)\right)\right] = 0 \quad (8)$$

where $u^{(HS)}(r)$ corresponds to the potential of effective rigid sphere with diameter $\sigma_{HS}$. The truncated Lennard-Jones-like particle is rigid at $T=0K$ same as $v \to \infty$. Two series of numerical calculation have been performed in this work, one will have a glass transition while cooling and one without, namely, the effective volume fraction $\phi_0 = 0.5$ and 0.6 at $T=0K$. The calculated $\sigma_{HS}$ and $\phi$ versus $T$ are plotted in Fig. 1. $\sigma_{HS}$ and $\phi$ are decreasing with increasing $T$. For $\phi_0 = 0.6$, it is greater than the critical volume fraction $\phi_c$ ($\phi_c \approx 0.563$) of glass transition for rigid particle [16], and it will get a glass transition while cooling; for $\phi_0 = 0.5$, no glass transition. In SCGLE theory, $F(k,t)$ and $F_s(k,t)$ cannot decay to zero with glass transition, so we only consider the temperature without glass transition for $\phi_0 = 0.6$.

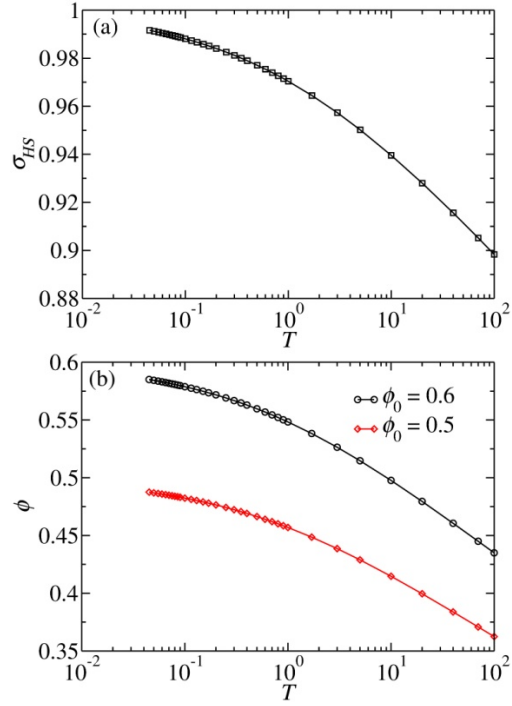

**Fig.1** The effective rigid sphere diameter $\sigma_{HS}$ and volume fraction $\phi$ as a function of temperature $T$: (a) $\sigma_{HS}$ and (b) $\phi$.

With the effective rigid sphere $\sigma_{HS}$ and volume fraction $\phi$, $D_L$, $\tau_s$ and $\tau_n$ were calculated; the related results are plotted in Fig. 2 and Fig. 3, respectively. $D_L$ is increasing with increasing $T$ for both $\phi_0 = 0.5$ and 0.6; which is larger for $\phi_0 = 0.5$ than that for $\phi_0 = 0.6$ at the same $T$. Both $\tau_s$ and $\tau_n$ are decreasing with increasing $T$ for $\phi_0 = 0.5$ and 0.6; besides $\phi_0 = 0.5$ relaxes faster than $\phi_0 = 0.6$ at the same $T$. The logarithms of $D_L$, $\tau_s$ and $\tau_n$ for $\phi_0 = 0.5$ and 0.6 are plotted in Fig. 4(a) and (b), respectively. The two variants of the Stokes-Einstein relation, $D \sim \tau^{-1}$ and $D \sim T/\tau$, were tested by $D \sim \tau_s^{\xi}$, $D \sim (\tau_s/T)^{\xi}$ and $D \sim (\tau_n/T)^{\xi}$, respectively. As debated above, the Stokes-Einstein relation is satisfied in SCGLE theory for it is an equilibrium theory; so if $\xi=-1$, the variant gives a consistent Stokes-Einstein relation and otherwise it is not a good variant.

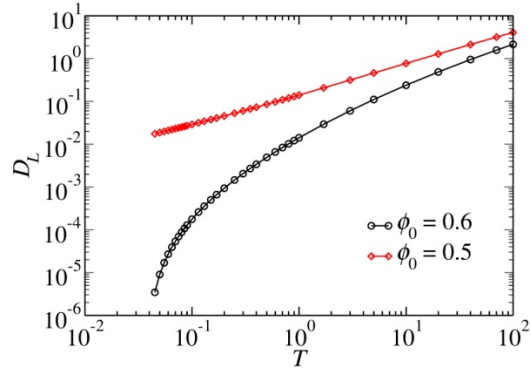

**Fig.2** The long time diffusion constant $D_L$ versus temperature $T$.

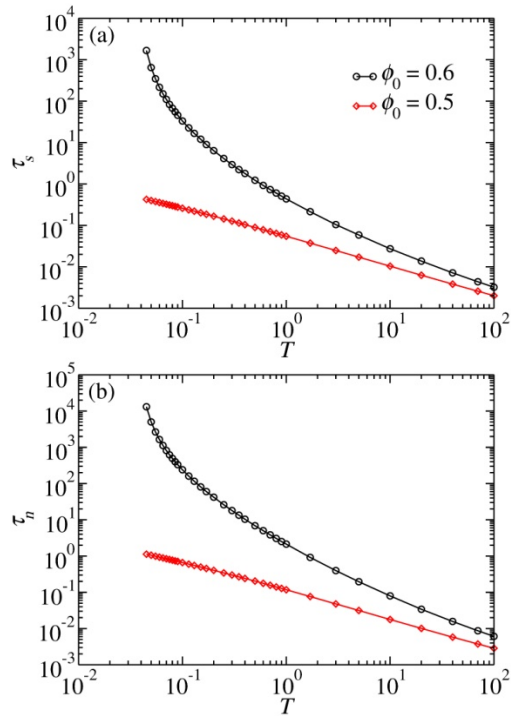

**Fig.3** The structural relaxation time as a function of temperature $T$: (a) $\tau_s$ described by $F_s(k,t)$; (b) $\tau_n$ described by $F(k,t)$.

It is shown that $D_L$ and $\tau_s$ can be well fitted by $D \sim \tau_s^{\xi}$ with $\xi \approx -1$ for both $\phi_0 = 0.5$ and 0.6; the results suggested that $D \sim \tau_s^{\xi}$ is a good variant of the Stokes-Einstein relation. In supercooled liquids, $D \sim \tau_s^{\xi}$ is shown a fractional form with $\xi$ greater than -1 due to the dynamic heterogeneity, the displacements of particle deviate from Gaussian [2, 9, 26]. The $D_L$, $\tau_s$ and $\tau_s$ for $\phi_0 = 0.5$ are also well fitted by $D \sim (\tau_s/T)^{\xi}$ and $D \sim (\tau_n/T)^{\xi}$; but $\xi$ is -0.42 or -0.4, which is not equal to -1. For $\phi_0 = 0.6$, the data needs two $D \sim (\tau_s/T)^{\xi}$ or $D \sim (\tau_n/T)^{\xi}$ with different $\xi$ to fit; a transition is observed while cooling. $\xi$ is different for the low $T$ and high $T$. $\xi$ is about -0.8 at lower $T$ and -0.5 at higher $T$. $\xi$ is closer to -1 for lower $T$ than that at higher $T$. The similar transition has also been observed in simulations, such as in water [6, 12]; but $\xi$ is almost equal to -1 at high $T$ and is larger than -1 at low $T$. The results suggested that $D \sim (\tau_s/T)^{\xi}$ and $D \sim (\tau_n/T)^{\xi}$ are not good variants of the Stokes-Einstein relation.

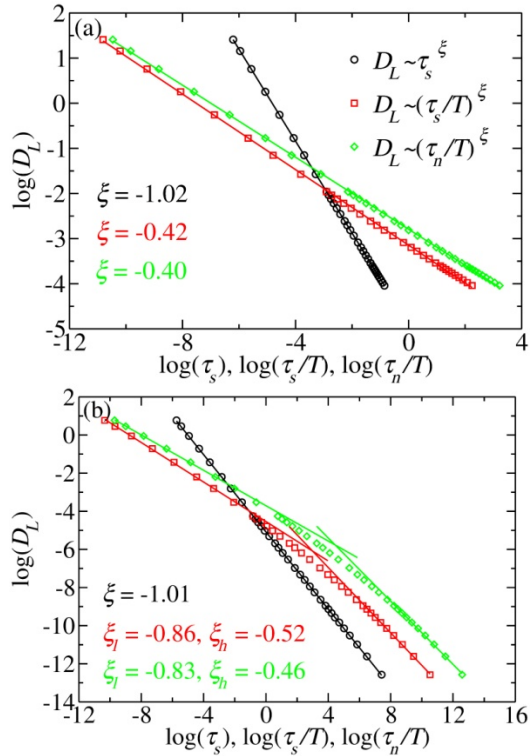

**Fig.4** The relation between the long time diffusion constant $D_L$ and the structural relaxation time $\tau_s$ or $\tau_n$: (a) $\phi_0 = 0.5$ and (b) $\phi_0 = 0.6$. The symbols are the numerical calculated data and the solid lines are fitted with different functional forms. The cycles, squares and diamonds are fitted by $D \sim \tau_s^\xi$, $D \sim (\tau_s/T)^\xi$ and $D \sim (\tau_n/T)^\xi$, respectively. The different colored $\xi$ corresponds to the same colored solid line. For $\phi_0 = 0.6$, the data need two lines in form $D \sim (\tau_s/T)^\xi$ or $D \sim (\tau_n/T)^\xi$ to fit, the subscript $h$ labels the higher temperature data and $l$ labels the lower temperature data.

It has been shown that $D \sim \tau^{-1}$ is a good variant but $D \sim T/\tau$ is not. To evaluate the shear viscosity $\eta$, $\eta$ is $\eta \sim T\tau$ for $D \sim \tau^{-1}$ and $\eta \sim \tau$ for $D \sim T/\tau$, respectively. Because of the dynamics equivalence between colloid and molecular liquids [18], $\eta$ can be evaluated by $\eta \sim T/D_L\sigma_{HS}$. Fig. 5 shows $\eta$ for $\phi_0 = 0.5$ and $\phi_0 = 0.6$ at different $T$. $\eta$ changes differently with $T$ for $\phi_0 = 0.5$ and $\phi_0 = 0.6$. $\eta$ is increased with increasing $T$ for $\phi_0 = 0.5$, but $\eta$ for $\phi_0 = 0.6$ is firstly fast decreased with increasing $T$ and then get slowly increasing with $T$. The two functional formal, $\eta \sim T\tau$ and $\eta \sim \tau$, were evaluated in Fig. 6. $\eta \sim T\tau_s$ is satisfied for $\phi_0 = 0.5$ and $\phi_0 = 0.6$; $\eta \sim T\tau_n$ is only satisfied for low $T$ and deviates at high $T$. $\eta \sim \tau_s$ and $\eta \sim \tau_n$ are awfully invalid that no any linear region is observed. Overall, $D \sim \tau_s^{-1}$ is a good variant of the Stokes-Einstein relation but $D \sim T/\tau$ is not a good variant in the framework of SCGLE theory.

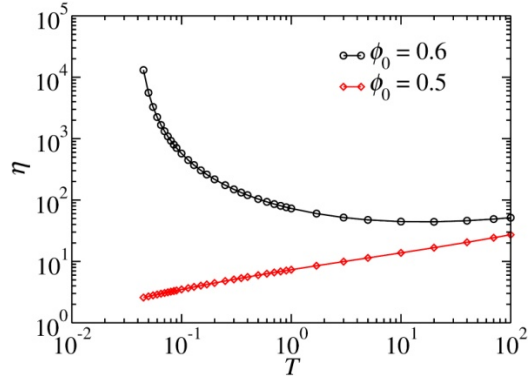

**Fig. 5** The shear viscosity $\eta$ versus $T$.

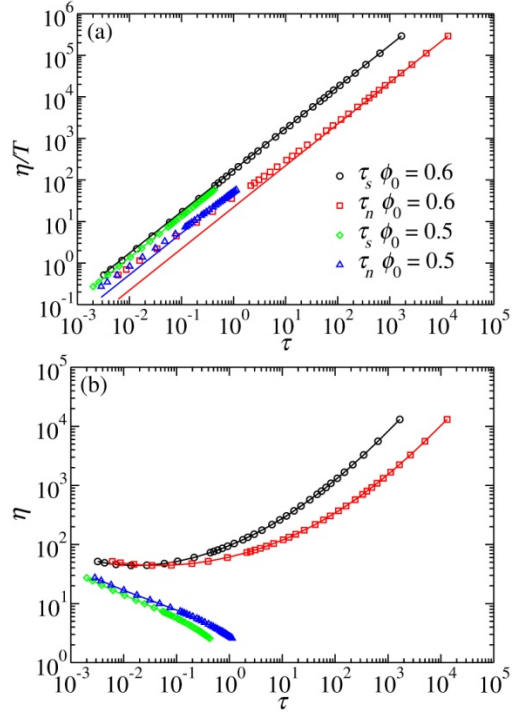

**Fig.6** The shear viscosity $\eta$ dependence on structural relaxation $\tau$: (a) $\eta/T \sim \tau$ and (b) $\eta \sim \tau$. The symbols are numerical calculated data, the solid lines in (a) are fitted lines with $\eta/T \sim \tau$.

## 4. Conclusion

In this work, taking two series of truncated Lennard-Jones-likes molecular liquids with effective volume fraction $\phi_0 = 0.5$ and 0.6 at $T=0$K, we examined two usually adopted variants of the Stokes-Einstein relation in supercooled liquid, namely $D \sim \tau^{-1}$ and $D \sim T/\tau$, in the framework of SCGLE theory. $\phi_0 = 0.6$ has a glass transition while cooling and $\phi_0 = 0.5$ without. The variant $D \sim \tau_s^{-1}$ is satisfied for both $\phi_0 = 0.5$ and 0.6 at all temperatures. Nonetheless, the functional $D \sim T/\tau$ separately expressed as $D \sim T/\tau_s$ and $D \sim T/\tau_s$ are both invalid for $\phi_0 = 0.5$ and 0.6 at all temperatures. For $\phi_0 = 0.5$, $D \sim T/\tau_s$ and $D \sim T/\tau_n$ are in a fractional form with $\xi$ around -0.4 for all temperatures. $D \sim T/\tau_s$ and $D \sim T/\tau_n$ display two fractional forms for $\phi_0 = 0.6$. A transition is observed while cooling and $\xi$ is jumped from a larger value to a smaller one. $\xi$ is about -0.8 at low temperature and -0.5 at high temperature.

Our numerical calculations are based on SCGLE theory with the structure factor calculated by the Percus-Yevick approximation complemented with the correction of Verlet and Weis. Our results indicate that $D \sim \tau_s^{-1}$ is a good variant of the Stokes-Einstein relation but $D \sim T/\tau$ is not. In simulations, $D \sim \tau_s^{-1}$ is observed to be invalid in supercooled liquids for the dynamic heterogeneity; $D \sim T/\tau$ also shows a transition while cooling, $\xi$ is closed to -1 at higher $T$ and deviates at lower $T$. In spite of the discrepancies between the ideal case predicted by SCGLE theory and simulations, the results in this work also give some evidences of the validity of the variants of the Stokes-Einstein relation; one needs to be careful while using $D \sim \tau^{-1}$ or $D \sim T/\tau$ to test the Stokes-Einstein relation.

**Acknowledgement**

This work was supported by the National Natural Science Foundation of China (No. 12104502) and the Science Foundation of Civil Aviation Flight University of China (No. J2021-054). The author Gan Ren thanks Pedro Ramírez-González (*Universidad Autónoma de San Luis Potosí*, *Mexico*) for suggestions.

# References

[1] R. Kubo, M. Toda and N. Hashitsume, *Statistical physics II: nonequilibrium statistical mechanics*. Springer Science & Business Media: 2012; Vol. 31.
[2] Z. Shi, P. G. Debenedetti and F. H. Stillinger, *J. Chem. Phys.* 138 (2013) 12A526.
[3] S. F. Swallen, P. A. Bonvallet, R. J. McMahon and M. D. Ediger, *Phys. Rev. Lett.* 90 (2003) 015901.
[4] M. K. Mapes, S. F. Swallen and M. D. Ediger, *J. Phys. Chem. B* 110 (2006) 507-511.
[5] F. Mallamace, M. Broccio, C. Corsaro, A. Faraone, U. Wanderlingh, L. Liu, C.-Y. Mou and S. H. Chen, *J. Chem. Phys.* 124 (2006) 161102.
[6] L. Xu, F. Mallamace, Z. Yan, F. W. Starr, S. V. Buldyrev and H. Eugene Stanley, *Nat Phys* 5 (2009) 565-569.
[7] K. Binder and W. Kob, *Glassy materials and disordered solids: An introduction to their statistical mechanics*. World Scientific: 2011.
[8] J. Habasaki, C. Leon and K. Ngai, *Top Appl Phys* 132 (2017).
[9] D. Jeong, M. Y. Choi, H. J. Kim and Y. Jung, *Phys. Chem. Chem. Phys.* 12 (2010) 2001-2010.
[10] L. O. Hedges, L. Maibaum, D. Chandler and J. P. Garrahan, *J. Chem. Phys.* 127 (2007) 211101.
[11] A. Ikeda and K. Miyazaki, *Phys. Rev. Lett.* 106 (2011) 015701.
[12] P. Kumar, S. V. Buldyrev, S. R. Becker, P. H. Poole, F. W. Starr and H. E. Stanley, *Proc. Natl. Acad. Sci. U. S. A.* 104 (2007) 9575-9579.
[13] F. de J. Guevara-Rodríguez and M. Medina-Noyola, *Phys. Rev. E* 68 (2003) 011405.
[14] P. E. Ramírez-González, L. López-Flores, H. Acuña-Campa and M. Medina-Noyola, *Phys. Rev. Lett.* 107 (2011) 155701.
[15] L. Yeomans-Reyna, M. A. Chávez-Rojo, P. E. Ramírez-González, R. Juárez-Maldonado, M. Chávez-Páez and M. Medina-Noyola, *Phys. Rev. E* 76 (2007) 041504.
[16] P. E. Ramírez-González and M. Medina-Noyola, *J. Phys.: Condens. Matter* 21 (2009) 075101.
[17] M. A. Chávez-Rojo and M. Medina-Noyola, *Phys. Rev. E* 72 (2005) 031107.
[18] L.-F. Leticia, M.-M. Patricia, E. S.-D. Luis, L. Y.-R. Laura, V.-R. Alejandro, P.-Á. Gabriel, C.-P. Martín and M.-N. Magdaleno, *EPL (Europhysics Letters)* 99 (2012) 46001.
[19] N. Xu, T. K. Haxton, A. J. Liu and S. R. Nagel, *Phys. Rev. Lett.* 103 (2009) 245701.
[20] M. Schmiedeberg, T. K. Haxton, S. R. Nagel and A. J. Liu, *EPL (Europhysics Letters)* 96 (2011) 36010.
[21] J. K. Percus and G. J. Yevick, *Physical Review* 110 (1958) 1-13.
[22] M. S. Wertheim, *Phys. Rev. Lett.* 10 (1963) 321-323.
[23] L. Verlet and J.-J. Weis, *Phys. Rev. A* 5 (1972) 939-952.
[24] R. Juárez-Maldonado, M. A. Chávez-Rojo, P. E. Ramírez-González, L. Yeomans-Reyna and M. Medina-Noyola, *Phys. Rev. E* 76 (2007) 062502.
[25] J. P. Hansen and I. R. McDonald, *Theory of Simple Liquids*. Elsevier Science: 1990.
[26] S. Kim, S.-W. Park and Y. Jung, *Phys. Chem. Chem. Phys.* 18 (2016) 6486-6497.